# Discrepant transport characteristics under Anderson localization at the two limits of disorder

**Authors**

Randhir Kumar[1], Sandip Mondal[1], M. Balasubrahmaniyam[1], Martin Kamp[2], and Sushil Mujumdar[1],

[1]Nano-optics and Mesoscopic Optics Laboratory, Tata Institute of Fundamental Research, Homi Bhabha Road, Mumbai, 400 005, India.

[2]Technische Physik, Universität Würzburg, 97074 Würzburg, Germany.

**Abstract**

Anderson localization is a striking phenomenon wherein transport of light is arrested due to the formation of disorder-induced resonances. Hitherto, Anderson localization has been demonstrated separately in two limits of disorder, namely, amorphous disorder and nearly-periodic disorder. However, transport properties in the two limits are yet unstudied, particularly in a statistically consistent manner. Here, we experimentally measure light transport across two-dimensional open mesoscopic structures, wherein the disorder systematically ranges from nearly-periodic to amorphous. We measure the generalized conductance, which quantifies the transport probability in the sample. Although localization was identified in both the limits, statistical measurements revealed a discrepant behavior in the generalized conductance fluctuations in the two disorder regimes. Under amorphous disorder, the generalized conductance remains below unity for any configuration of the disorder, attesting to the arrested nature of transport. Contrarily, at near-periodic disorder, the distribution of generalized conductance is heavy-tailed towards large conductance values, indicating that the overall transport is delocalized. Theoretical results from a model based on the tight-binding approximation, augmented to include open boundaries, are in excellent agreement with experiments, and also endorse the results over much larger ensembles. These results quantify the differences in the two disorder regimes, and advance the studies of disordered systems into actual consequences of Anderson localization in light transport.

**Introduction**

Wave transport in mesoscopic systems is an area of ubiquitous interest [1, 2]. Mesoscopic fluctuations essentially arise through self-interference of scattered waves, leading to weak localization and Anderson localization among other effects [3-6]. In recent years, the topic of Anderson localization has transcended the boundaries of condensed matter physics into other domains such as optics, acoustics, quantum matter etc. [7-10] Among these, light localization has made massive strides with the discovery of some exotic phenomena involving nonlinearity [11, 12], quantum correlations [13], hyper-transport [14], correlated scattering [15], photonic Anderson insulators [16], lasing [17-21], and so on. On one hand, the earliest observations of weak localization of light in amorphous disorder [22,23] motivated the usage of strong amorphous disorder for Anderson localization [9, 17,18, 24-27]. On the other hand, seminal theoretical works suggested an interesting route to photon localization in nearly-periodic systems [28], which has been exploited in several low-dimensional systems [21, 29-33]. While both the routes to localization are vastly successful, the similarities and differences in the physics of transport in these two diverse scenarios need to be carefully investigated.

Experiments on 3D Anderson localization focused on transport parameters such as transmission [24], temporal pulse expansion [25] etc. towards identification of localization. On the other hand, experiments in two-dimensional systems [11, 26, 32, 33, 36-39] took advantage of direct access to the eigenmodes. For instance, a pioneering set of experiments directly traced the transition from diffusive states to Anderson localized states in a photo-induced transversely disordered structure, which mapped the eigenstates of the ballistic/diffusive/localized light [11]. In a two-dimensional membrane structure embedded with emitters, localized eigenmodes were identified in near-periodic structures to measure non-universal correlations [32]. More recently, optical structures were realized that simulated disordered potentials in crystalline lattices, wherein Anderson localized eigenmodes were imaged in the band-tails [33]. While measurement of the eigenmodes is a crucial step towards identifying localization, actual transport through the system involves further factors. For instance, a quasiparticle whose energy does not match an eigenfrequency can still experience nonresonant transport in an open finite system. Similarly, in an open periodic structure, the density of states does not vanish in the stopbands due to the coupling to the external electromagnetic vacuum, thus allowing transport in the stopband [40,41]. The twin conditions of finite size and coupling to the environment, known in short as finite support [42], play a vital part in the physics of transport in open mesoscopic systems.

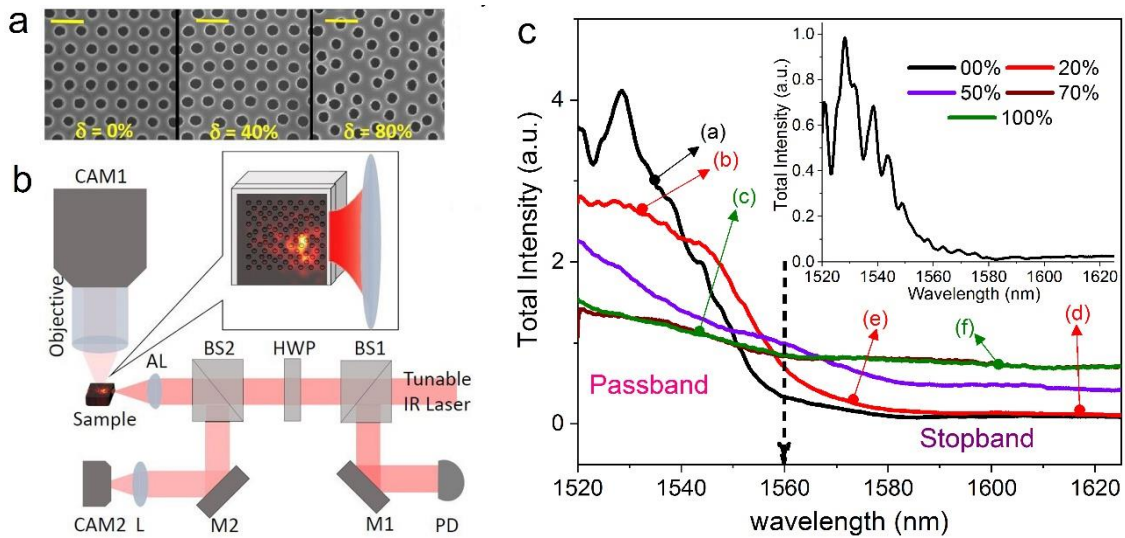

**Fig. 1. Experimental measurement and characterization.** (a) SEM images of periodic, $40\%$ disordered and $80\%$ disordered sample. Scale bar $1\mu m$. (b) Experimental setup for measurements. (See method section for details.) (c) Configurationally-averaged admittance into the membrane for various disorder strengths. The lower bandedge (dashed arrow at $\lambda \sim 1569nm$) of the system can be identified in the periodic situation (black curve). The bandedge is washed out with increasing disorder. Intensity distributions in some configurations at the labeled points are illustrated later. Inset: Admittance of light in an individual configuration in the periodic system. Systemic modes can be identified that are broadened due to the finite support to allow nonresonant transport.

In this work, we address the physics of overall transport under conditions of Anderson localization at either limit of disorder in open systems. We realize a passive membrane-based disordered system wherein the transport is studied by launching monochromatic light into one open boundary of the sample. The disorder is varied over a large range, starting from periodic structure to perturbed-periodic and finally to strongly disordered amorphous structures. Band-resolved measurements allow us to measure Bloch eigenmodes in the periodic structure, and Anderson localized modes in strong disorder and also at the bandedge in nearly-periodic disorder. We characterize the transport through generalized conductance fluctuations, and observe that they differ significantly under localization at the two limits of disorder. We implement a model based on the tight-binding approximation modified to allow for finite support. The computed results are in excellent agreement with the experimental observations.

**Results.**

Figure 1(a) depicts the SEM images of a section of the photonic crystal samples at three different disorder strengths (defined in the Methods section) $\delta = 20\%, 40\%, 80\%$, respectively. The scale bar is $1\mu m$. (b) shows the experimental setup used in this study, and the detailed procedure is elaborated in the Methods section. Laser light from a tunable diode laser was edge-coupled to the samples, and transport measurements were made. As the wavelength of the incident light was tuned, the periodic/disordered membrane admitted the light depending upon the available modes. The out-of-plane scatter from the membrane surface represents the intensity distribution inside the membrane [31]. Accordingly, the admittance of the membrane is quantified by first imaging the out-of-plane scattered intensity on a CCD, and then spatially-integrating the image to obtain a single intensity value at a given wavelength. Figure 1(c) shows the configurationally-averaged admittance as a function of wavelength for the various disorder strengths. Each profile corresponds to a particular disorder strength, labeled by the color in the legend. In the periodic sample at $\delta = 0\%$ (black line), a rapid fall of the admittance reveals the bandedge at $\lambda = 1560nm$ (black dotted arrow). Although the exact bandedge could not be identified from the spectra, we note here that a shift of $\pm 5nm$ did not affect the analysis nor our conclusions. Particularly noteworthy is the profile at $20\%$ disorder, where a visible tail into the bandgap is seen, which corresponds to the Lifshitz tail in the density of states. With increasing disorder, the tail flattens out and the memory of the bands is completely lost. Since the periodic structure also involves unavoidable fabricational disorder, we illustrate the admittance of single periodic configuration in the inset. The finite size of the sample resolves the field into modes that can be identified as peaks in the passband region. Clearly, the peaks are broadened to about $10nm$, and merge with each other. At the wavelengths between peaks, there is a finite conductance that signifies nonresonant transport. In the stopband region, the amplitude is small, but nonzero, implying very small conductance. In the main plot, representative points on either side of the bandedge are chosen that describe the typical mode structure as it evolves with disorder.

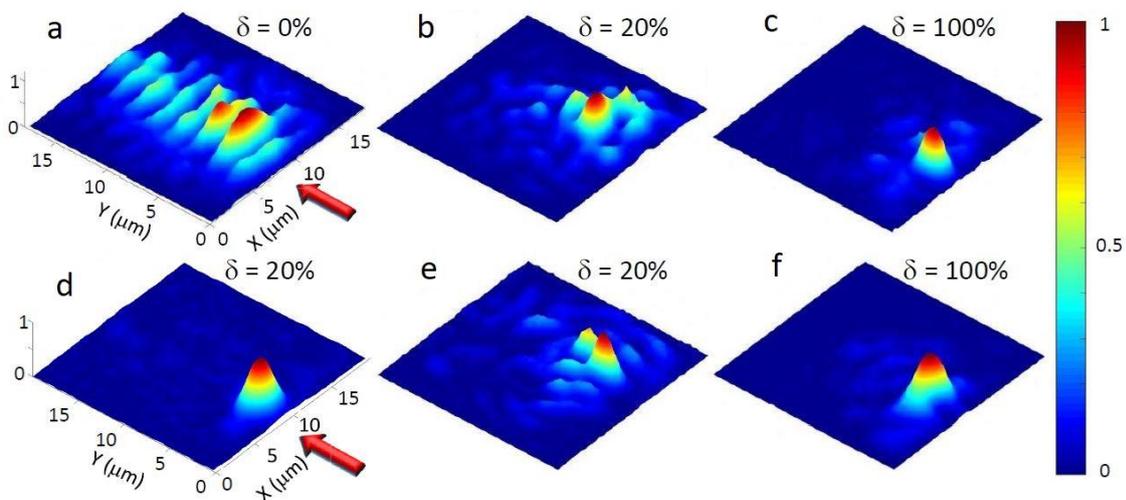

Fig. 2. **Experimental Mode Profiles.** Intensity distributions at the wavelengths labeled in Fig1. Upper panel represents modes in the passband range, while lower panel depicts modes measured in the stopband range. [a] Photonic mode seen in the periodic system in the passband ($1534.8nm$). [b] and [e] are weakly localized modes at $20\%$ disorder in the passband ($1532.2nm$) and stopband ($1573.3nm$) respectively. [c] Strongly localized mode at passband wavelength ($1543.5nm$). [d] Strongly localized mode in the stopband ($1617.0nm$) even at weak disorder of $20\%$. [f] Strongly localized mode at the stopband wavelength ($1601.0nm$). At high disorder ($\delta > 50\%$), there are no band effects.

Figure 2 describes these mode structures at various points marked. The red arrow in two images indicates the input edge of the membranes, consistent with (but not shown for) all the images. The evolution of the intensity distribution in the passband is described by the upper panels. In the periodic sample (Fig. 2(a)), parallel wavefronts comprising a superposition of counter-propagating Bloch modes are seen. With introduction of disorder, the light undergoes weak localization as seen at $20\%$ disorder (Fig. 2(b)). In the limit of strong disorder, all the light field is strongly concentrated to provide Anderson localization. In the bandgap wavelengths (lower panels), Anderson localization is immediately achieved even at $20\%$ disorder in the vicinity of the bandedge, as seen in Fig. 2(d). On the other hand, Fig 2(e) shows an image at another wavelength, also at $20\%$ disorder, where the light is only weakly localized, typifying the statistical nature of the localization at this disorder. Finally, Fig. 2(f) shows another localized mode at $100\%$ disorder.

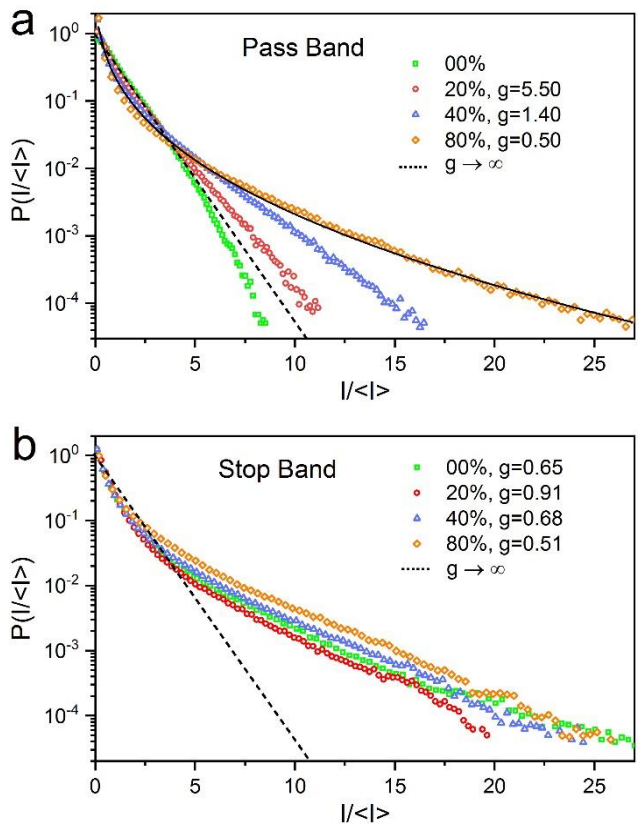

**Fig. 3. Intensity distribution of modes.** (a) $P(I/\langle I \rangle)$ in the passband ($\lambda = 1520 - 1560nm$) for four disorder strengths, ranging from periodic to strong disorder. The solid line shows the bestfit for the highest disorder. The dashed line indicates $g \to \infty$. (b) $P(I/\langle I \rangle)$ for modes in the stopband ($\lambda = 1560 - 1625nm$), with the bestfit . All values indicate localized transport.

While the images endorsed Anderson localization in the two limits of disorder, the quantification of transport over the entire range of disorder was carried out using the conductance $g$. It is known from theory that the distribution of intensity fluctuations shows a well-defined analytic behavior, which enables a bestfit characterization of the conductance $g$ [5, 32, 45] (See Methods section for details). Accordingly, the conductance $g$ was first computed from the intensity distribution realized over all configurations and wavelengths, and studied separately over the passband and the stopband. Figure 3 depicts the measured distribution in the experiments, with subplot (a) showing the behavior in the passband. For the strongest disorder (80%), a clear long tail is observed. The bestfit shown with a solid line yields the $g = 0.5$, quantifying the localization in the limit of strong disorder. With reduced disorder, the tails are clearly shorter, and the fits (not shown to maintain clarity) yield $g$ larger than 1, as described in the legend. This is the domain of renormalized transport, where the light is not localized but is slower than diffusive. The dashed line demarcates an exponential decay of the Rayleigh distribution, which represents diffusive transport where $g \to \infty$. Clearly, the periodic system (green markers) exhibits a slower-than-exponential trace, and represents quasi/ballistic transport. This variation cannot be fit by the theory and hence the $g$ cannot be quantified. Next, Fig. 3(b) depicts the measurements in the stopband. In contrast to the passband behavior, the $P(I/\langle I \rangle)$ show a long-tailed behavior for all disorders. At 20% disorder, the was found to be 0.91, which indicates localized transport in the bandgap range, i.e., in the limit of near-periodic disorder. One interesting observation is in the behavior in the periodic sample, which possesses a small inherent fabricational disorder. Despite the lack of existence of a mode in the system, the $P(I/\langle I \rangle)$ is not arbitrary, but on the contrary, it follows a similar long-tailed curve as in the other disorder strengths. Clearly, all fits yield $g < 1$, a value indicating that the transport remains in the localization regime. Interestingly, $g_{\delta=0\%} < g_{\delta=20\%} > g_{\delta=40\%}$. This indicates an enhanced transport with the introduction of disorder into the periodic system, after which the conductance again decreases. It is known that the gap states formed due to disorder allow for transport in the erstwhile bandgap where no transport is envisaged. The conductance depends upon the number of gap states excited and their spatial localization, which is reflected in the $P(I/\langle I \rangle)$. The theoretical fit only offers a single value for $g$, which does not reflect the configurational fluctuations in the modes seen in the experiments. The theoretical expression can't be used for individual configurations as the limited sample size precludes the manifestation of the long tail in $P(I/\langle I \rangle)$. Therefore, towards a first approximation of configurational statistics, we adopted the technique that exploits the variance of intensity fluctuations.

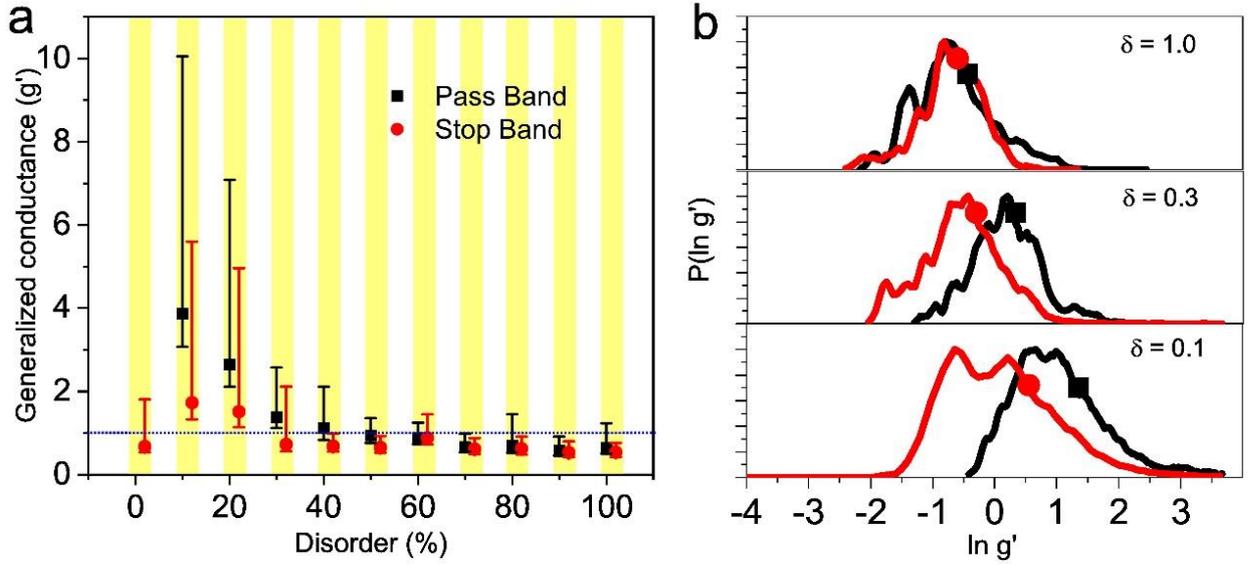

**Fig. 4. Experimental Conductance.** (a) Generalized conductance fluctuations ($g'$) as a function of disorder. Passband (black markers), and stopband (red markers) data are horizontally offset for clarity. Markers represent the mean, while the error bars depict the standard deviation in the fluctuations, with the asymmetry reproducing the asymmetry in the distribution. The fluctuations in $g'$ reduce with increasing disorder. The $\langle g' \rangle$ in the passband decreases monotonically and enters the localization regime at $\delta = 50\%$. The stopband $\langle g' \rangle$ initially increases at introduction of disorder, and subsequently decreases and then follows the passband $\langle g' \rangle$. (b) Three representative distributions of $\ln g'$ indicating the progression from an asymmetric distribution to a Gaussian distribution with increasing disorder. Markers indicate the mean values.

If $S_{ab}$ is the total transmission normalized to the ensemble average $S_{ab} = T_{ab}/\langle T_{ab} \rangle$, the generalized conductance $g'$ is given by $4/\{3(var(S_{ab}) - 1)\}$ [27]. The condition $g' < 1$ identifies localization. Accordingly, the variance of measured intensity distributions enabled us to compute $g'$ for all configurations in the wide range of disorder. (See Supplementary Information, section S2, page 20 of this document.) Towards an estimate of the fluctuations in $g'$, the variance of the intensity distribution at each wavelength was calculated. Next, statistics was carried out over the various configurations and wavelengths in the passband and the stopband. The resulting trends in the generalized conductance fluctuations are plotted in Fig. 4(a), with the mean depicted by the markers, and the error bars indicating the standard deviation in the $g'$. In the passband, at $\delta = 0\%$, the variance drops below 1 and hence the $g'$ is not defined. This is consistent with the bestfit technique for $g$, where $g$ cannot be fit due to the slower-than-exponential decay (Fig 3). The maximum $\langle g' \rangle$ was measured at the minimum disorder $\delta = 10\%$, after which it steadily decreases with disorder. The $\langle g' \rangle$ drops below 1 at $\delta = 40\%$. The fluctuations in $g'$ also reduce with increasing disorder. In the stopband, for the periodic sample, the minimal fabrication disorder in the sample realizes a spread in $g'$. There is a rise in $\langle g' \rangle$ at weak disorder followed by a steady decrease. This trend is in agreement with the behavior of $g$ in Fig 3. The fluctuations in $g'$ reveal the statistical behavior of the system. At weak disorder, while there could be some configurations for which $g' < 1$, there are several other configurations wherein the field is delocalized. Close to the $\delta = 50\%$, the two curves behave in a similar fashion, endorsing the loss of band-dependence. The asymmetric errorbars reflect the asymmetry in the $P(g')$. Figure 4(b) shows the $P(\ln g')$ for three representative disorders. It shows that $\ln g'$ at weak disorder has an asymmetric distribution with a long tail towards the high $\ln g'$. This implies that the propensity of modes at weak disorder is to remain delocalized, while only the rare modes undergo localization. At stronger disorder, however, $P(\ln g')$ approaches a normal distribution, and the majority of the modes are localized.

The experimentally measured $g'$ reveal that the localization properties at the two limits of disorder differ in the statistics. The localized modes seen at weak disorder in the stopband are occasional modes within the errorbar while the average transport is delocalized. In comparison, the localization that happens at strong disorder holds true also for the average transport.

Next, we proceeded to theoretically model the transport in a bid to understand the spectrally-resolved $g'$ over the wide range of disorder. The experimental observations necessitate a technique that can invoke nonresonant transport, and also allows for averaging over numerous configurations. Although time-domain techniques such as finite-difference time-domain simulations can compute nonresonant transport, they are too computationally-expensive when multiple configurations and/or large system sizes are involved. We, therefore, used the tight-binding approximation, which is extensively used in the electronic domain and identifies the eigenvalues and eigenmodes of a periodic/disordered system. We, then, augmented the method to compute nonresonant transport to simulate the experimental structures. The computed intensity distributions confirmed their utility for this analysis. We claim that this modified technique provides a viable cost-effective alternative to the FDTD method in disordered systems.

The theoretical formulation was initiated using the tight-binding formulation with diagonal disorder for a two-dimensional triangular lattice. To this end, we employed a tight-binding Hamiltonian described as,

$$H_{dc} = \sum_{i=1}^{M \times N} [\Omega_i c_i^\dagger c_i + \sum_j p c_i^\dagger c_{j+1} + h.c.] \tag{1}$$

where the $c_i$ is the annihilation operator of the fundamental mode, $p$ is the hopping probability between the sites, and $j$ runs over the nearest neighbours. The diagonal term is set to a uniformly distributed random number $\Omega_i \in [1 - W, 1 + W]$ where $W$ varies from 0 for a periodic system to 0.24 for a highly disordered system. The hopping probability $p$ is kept constant at 0.1. The Hamiltonian matrix is diagonalized to find the eigenvalues and hence the density of states. This transmission band ranged from $E = 0.7$ to $1.6$. (See Supplementary Information, section S3, page 21 of this document.) Different system sizes $M \times N$ are implemented. One thousand configurations are computed for statistical completeness at each disorder, and at each system size.

Figure 5 illustrates the results for one configuration at weak disorder $(W = 0.04)$ for a structure of $20 \times 34$ lattice sites. The graphic in Fig. 5(a) shows the computed eigenvalue spectrum in the vicinity of the lower bandedge, which was identified from the periodic structure and is marked by the vertical red line. The lower bandedge is chosen for analysis towards consistency with the experiments. The finite size of the structure resolves a continuous band into a series of $\delta$-functions at the eigenfrequencies. A range of wavelengths depicted by the vertical blue lines in Fig. 5(a) was chosen for averaging, both in the passband and the stopband regions. The weak disorder realizes the migration of the bandedge modes into the stopband as seen in the plot. For all eigenvalues $E_i$, the associated eigenvectors $\psi_i$ were computed.

Next, finite support was invoked by modeling it as loss to the boundary, quantified by a loss term $\Gamma$. Under occurrence of localization, the localization length $\xi$ determines the coupling to the environment [36], but $\xi$ can be accurately estimated only for strong disorder. Since this analysis applies over the wide range of disorder ranging from periodic to strong, a generic scaling length is needed to be identified. Hence, we invoke the inverse participation ratio, which is the second moment $q_2$ of the spatial integral of the density of the eigenfunction. In terms of intensity, $q_2 = N \sum_{i=1}^{N} I_i^2 / (\sum_{i=1}^{N} I_i)^2$, where $N$ is the number of spatial points over which the mode is computed. The $q_2$ quantifies the physical extent of the mode and, therefore, its coupling to the boundaries. Given a crystal size of $L$, we set $\Gamma \propto exp(-2R\sqrt{q_2}/L)$, where $R$ quantified the distance of the peak of the $\psi_i^2$ from the boundary. Note that four values of $\Gamma$ were initially calculated corresponding to each boundary, and the average value was ultimately used. The term $L/\sqrt{q_2}$ emulates the decay length of the mode. At high disorder, the term behaves as the localization length $\xi$, while at low disorder where $\xi$ cannot be estimated, the term provides a corresponding decay length. (See Supplementary Information, section S4, page 22 of this document.)

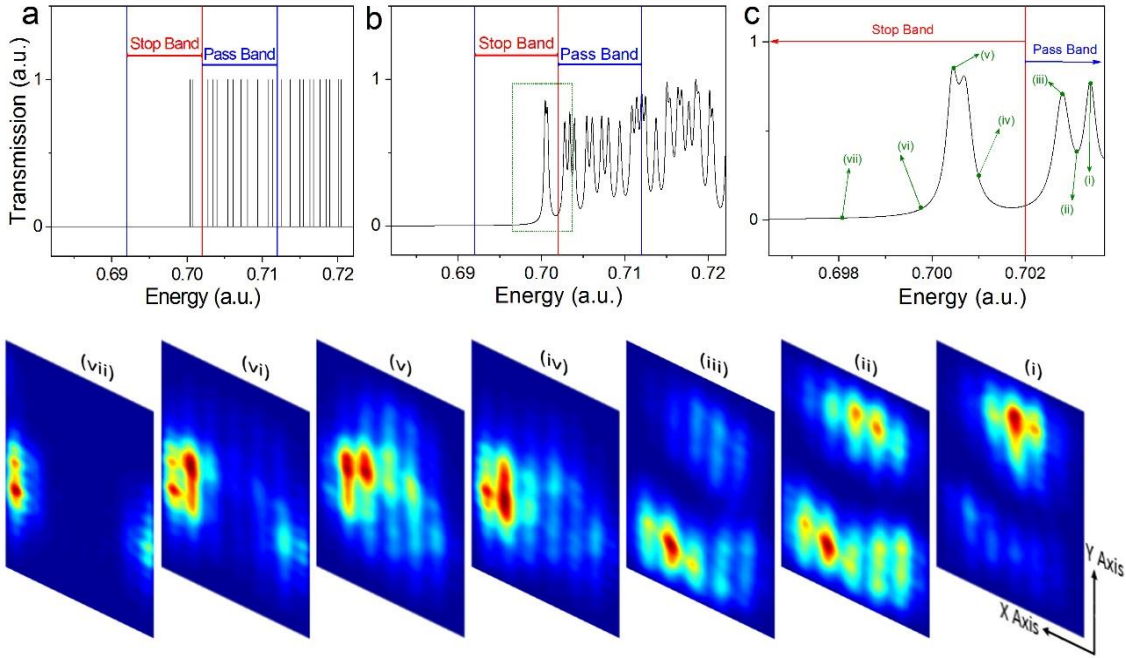

**Fig. 5. Theoretical Modes.** Computed intensity distributions from the model. (a) Computed eigenvalue spectrum for a weakly disordered $W = 0.04$ structure. Vertical red line demarcates the bandedge, blue lines identify energy ranges used for averaging. (b) Corresponding spectrum after introduction of boundary losses, as computed from the coupling of the modes to the environment. Dotted green rectangle is emphasized in (c). Seven wavelengths are chosen from the plot in (c) for illustration of the intensity distribution at various wavelengths, shown in the lower panel.

Accordingly, every $E_i$ was associated with a corresponding loss term $\Gamma_i$, which led to the broadening of the $\delta$-functions into Lorentzian profiles $\mathcal{L}_i$ whose sum provides the spectrum of the structure. (See Supplementary Information, Section S5, page 23 of this document, for periodic and moderate disorder.) Figure 5(b) depicts the spectrum of (a) after invoking the finite-support loss. The broadening of $E_i$ introduces a finite and substantially high transmission at several nonresonant wavelengths, as seen from the amplitudes between the peaks in the spectrum. The green dotted rectangle is emphasized in (c), in which seven representative wavelengths, as marked by the letters, are chosen for later discussion.

The reconstruction of the intensity distribution at all wavelengths is implemented as elaborated in the bottom panel of Fig 5. At a wavelength $\lambda$, the intensity distribution is described by,

$$I(x,y,\lambda) = \left| \mathcal{L}_i(\lambda) \psi_i(x,y) \exp \sum \left( \frac{\frac{2x}{L}}{\ln \mathcal{L}_i(\lambda)} \right) \right|^2 \qquad (2)$$

Here, $\psi_i$ is the eigenvector of $i^{th}$ eigenvalue. $\mathcal{L}_i(\lambda)$ represents the contribution of the $i^{th}$ eigenvector at $\lambda$, which would be zero in the absence of support. Apart from the eigenmodes, the vacuum outside the finite system also couples into the structure, resulting in exponentially decaying fields [40,41], which is represented by the exponential term in the Equation 2 [46]. For wavelengths far away from the eigenvalues, such as in the bandgap or between two eigenvalues, the small $\mathcal{L}_i(\lambda)$ ensures a tight decay. On the other hand, for wavelengths where modes exist, the exponential decay is very gradual due to a large $\mathcal{L}_i(\lambda)$, and essentially scales $\psi_i$ by a constant. In Fig 5(c), points (i), (iii) and (v) are at the eigenvalues of the Hamiltonian, and their intensity distributions are closest to the original eigenvectors. The points (ii), (iv), and (vi) represent the nonresonant transmission at the respective wavelengths. As can be seen, the intensity distribution at the point (ii) is seen to be derived from a superposition of $\psi$ at (i) and (iii). There are contributions from the other eigenvectors which are not evident due to their weak amplitudes at $\lambda_{ii}$. The point (vii) is quite far from any of the eigenvalues, due to which the intensity distribution primarily consists of the incoupled vacuum fields. Thus, the resonant and nonresonant intensity distributions are obtained in the model. Excellent agreement was obtained in the experimentally-measured and the theoretically computed intensity distributions. (See Supplementary Information, section S6, page 24 of this document.) Furthermore, the local density of states (LDOS) at the center of the sample in the bandgap was found to decrease exponentially with increasing sample size, as is expected with finite support [41]. (See Supplementary Information, section S7, page 25 of this document.)

Having thus computed the intensity distributions in finite supported structures over the wide range of disorder, we calculated the generalized conductance fluctuations $g'$. Figure 6(a) depicts the computed $g'$ with the markers representing $\langle g' \rangle$ and the errorbar signifying the standard deviation of $g'$. With increasing disorder, in the passband (black markers), the $\langle g' \rangle$ decreases and traverses below 1 at $W = 0.12$. At $W = 0$, the variance is smaller than 1, and hence the $g'$ is not defined. In the stopband (red markers), the $\langle g' \rangle$ shows a non-monotonic behavior, which initially increases with $W$, and subsequently starts reducing again. The peak conductance in this structure occurs at $W = 0.04$. It follows the variation for the passband at $W = 0.14$. Evidently, the band-dependence is lost by this disorder strength. The errorbar shows the asymmetric distribution about the mean. The fluctuations in $g'$ are also seen to decrease with increasing disorder. Three representative distributions of $\ln g'$ are shown in Fig 6(b), with the mean values marked explicitly. Clearly, at strong disorder, the distribution of $\ln g'$ tends to a Gaussian. In comparison, at near-periodic disorder in the stoband, a heavy-tailed asymmetric distribution of the generalized conductance occurs. (See Supplementary Information, section S8, page 26 of this document, for the complete set of $P(\ln g')$.) These theoretical results are in excellent agreement with the experimental observations. Similar to the experiments, the averaging was carried out over a range of energies, $\Delta E = 0.006$ specifically. However, the availability of arbitrary number of configurations in the simulations motivated us to carry out the averaging over narrower range of wavelengths. Figure 6(c) and (d) illustrate the $P(\ln g')$ in the near-periodic disorder and the amorphous disorder over a range $\Delta E$ of 0.003, while Figure 6(e) and (f) show the same over $\Delta E = 0.0015$. This proves that the fluctuations in $g'$ at weak disorder are larger and asymmetrically distributed than that at strong disorder, regardless of the spectral bandwidth chosen for the statistics.

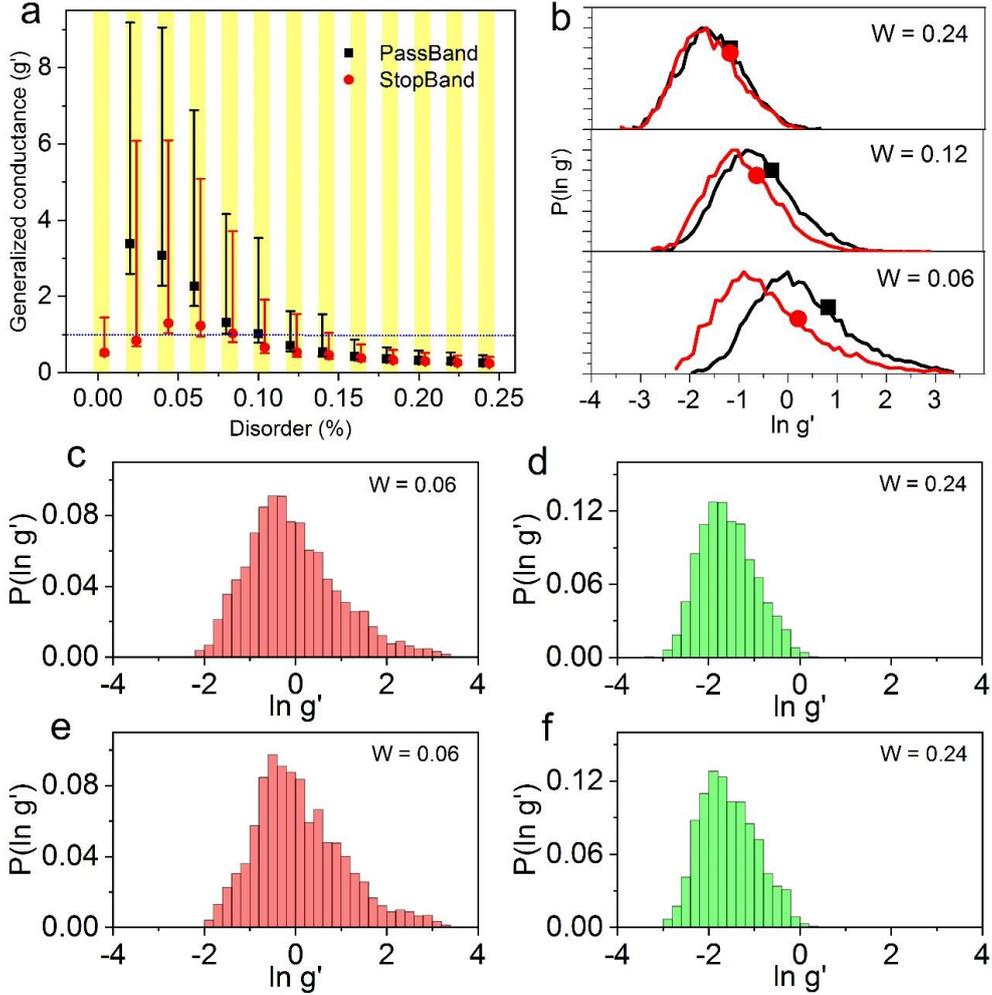

**Fig. 6. Theoretical Conductance.** Computed generalized conductance $g'$ from the model. (a) Computed $g'$ from the variance of $|\psi|^2$ for the passband (black markers) and the stopband (red markers). Markers show the mean, and the error bars show the standard deviation in $g'$. In the passband, the $\langle g' \rangle$ reduces with disorder, while in the stopband, it initially rises before decreasing with disorder. (b) $P(\ln g')$ for three values of $W$. The computed behavior follows the experimental observations excellently. Computed behavior for $g'$ averaged over smaller energy ranges, namely $\Delta E = 0.003$ in the stopband at weak-disorder (c) and at strong disorder (d), and $\Delta E = 0.0015$ in the stopband at weak-disorder (e) and at strong disorder (f).

Furthermore, the trend in the results remains the same with system size, and the magnitudes of $g'$ reduce with increasing size. (See Supplementary Information, section S9, page 27 of this document.) The transition from delocalized to localized modes occurs at weaker disorder for larger samples in the passband. With increasing sample size, the loss parameter reduces, and so the fall in $\langle g' \rangle$ with disorder is faster. In the stopband, the $\langle g' \rangle$ can remain below 1 even for weak disorder at larger sample size for strong disorder.

Interestingly, both the experiments and theory show that the $\langle g' \rangle$ does not actually represent the average behavior at the corresponding disorder strength, a fact which is generally associated with $\langle g \rangle$, the conductance, in the literature of disordered systems. At weak disorder, the strong asymmetry offsets the mean to very large values. At strong disorder, the $\langle \ln g' \rangle$ reasonably describes the logarithmic fluctuations, but still the $\langle g' \rangle$ is significantly skewed to the larger values.

**Conclusion**

In conclusion, we have experimentally studied Anderson localization in two-dimensional, open, mesoscopic systems, wherein localization was achieved by both the known routes, at the bandedge region in nearly-periodic systems, and in the presence of strong disorder. The openness of the system was necessary for the light injection into the system. The probability density of the transported quasiparticles (in this case, photons) was measured in the form of intensity distributions over several wavelengths and disorder configurations. The generalized conductance $g'$ essentially quantifies the probability of transport. A discrepant behavior was recorded in the localization properties at the two limits of disorder when statistical ensembles of $g'$ were analyzed. In the presence of amorphous disorder, the $\ln g'$ had a Gaussian distribution with almost all values being less than zero, indicating strong localization. In the near-periodic disorder in the stopband, the $\ln g'$ exhibited an asymmetric heavy-tailed distribution with a large number of values lying above zero. This shows explicitly that average transport will not be arrested under Anderson localization conditions in near-periodic disorder, although localized modes do exist. On the other hand, under amorphous disorder, the system will arrest all transport. Extrapolating these observations to three-dimensions, it essentially underlines that the metal-insulator transition arising from Anderson localization will occur only under conditions of strong disorder. This observation is significant because, the literature on electronic localization in three dimensions always refers to strong disorder. Indeed, localized electrons under weak disorder are referred to as 'trivially localized', implying the assistance of reduced density of states in the bandgap. In the photonic scenario, near-periodic disorder makes a strong case if the aim is the observation of Anderson localization of the band-tail states.

This study sheds a significant light on the physics of wave transport in open disordered structures, beyond the mere observation of localization. We believe it will enable further research activity in this rich and exciting field.

**Materials and Methods**

Design of disorder: The primary periodic structure was designed using the MIT Photonic Bandgaps (MPB) freeware, which yielded the bandstructure for a hexagonal air-hole lattice (lattice constant $a = 629nm$, hole radius $r = 138nm$) in a GaAs membrane such that the bandedge lay in the range of $1520nm - 1625nm$, coinciding with an existing diode-laser source. Disordered samples were designed by displacing the center $(x, y)$ of each air-hole by uniformly distributed random numbers $(\partial x, \partial y)$, such that $\partial x^2 + \partial y^2 \leq d^2$, where $d = \frac{M}{2} \times \frac{\delta}{100}$. Here, $M = a - 2r$ is the maximum possible displacement to avoid coalescence of two holes. The disorder parameter $\delta$ was varied from a disorder of $0\%$ (periodic) to $100\%$ implying complete disorder, in steps of $\delta = 10\%$, realizing structures from periodic, to periodic-on-average random, to amorphous. Twenty-five configurations at each disorder strength were designed and fabricated. The membrane thickness was $340nm$, while the region of the airholes was of lateral dimensions $20\mu m \times 20\mu m$.

Sample Fabrication: Sample fabrication was carried out in four stages, as per established techniques [43, 44]. (I) A $339nm$ thick GaAs layer was grown on a thick sacrificial Al$_{0.7}$Ga$_{0.3}$As layer in a solid source molecular beam epitaxy system on a (001) GaAs wafer. After the epitaxy, a $100nm$ thick SiO$_2$ layer is deposited on the sample by RF-magnetron sputtering. Then, a $500nm$ thick layer of electron beam resist, Polymethyl methacrylate (PMMA) with 950 k molecular weight, is spin coated on the sample. The e-beam lithography of the disordered photonic crystal patterns is carried out with an electron energy of 100 keV with 2 nA beam current. The sample is then developed in a 1:3 mixture of Methyl-isobutyl-ketone (MIBK) and isopropanol (IPA) for 2:30 min, followed by a 15s rinse in pure IPA, which removes the exposed parts of the resist. A schematic of the same is shown in the Supplementary document, Section S1, page 19 of this document. (Supplementary Fig. S1(a)). (II) The pattern is then transferred into the underlying SiO$_2$ layer using a reactive ion etch step with a CHF$_3$/Ar chemistry (flow rates: 15 sccm CHF$_3$ and 7:5 sccm Ar, pressure: $2.3 \times 10^{-3}$ mbar, radio frequency power: 50 W, etch time: 20 min). (Supplementary Fig. S1(b)). (III) After removal of the remaining PMMA, the pattern is etched further down into the semiconductor using an electron cyclotron resonance reactive ion etch step (ECR RIE) with a Cl$_2$/Ar mixture (flow rates: 3:5 sccm Cl$_2$ and 27 sccm Ar, pressure: $3 \times 10^{-3}$ mbar, RF power: 70 W, ECR power: 250 W, 8 min etch time). (Supplementary Fig. S1(c)). (IV) The membrane is now under-etched (etch time: 25-30s) using diluted HF. This also removes the remaining SiO$_2$ layer. This results in the final structure. (Supplementary Fig. S1(d).)

Experimental measurement technique: The experimental setup is schematized in Fig. 1(b). A collimated NIR laser beam from a wavelength tunable ($\lambda = 1520nm - 1625nm$) diode laser was passed through a combination of a polarising beam splitter (PBS) and a half wave plate (HWP) to obtain the desired polarization. The beam was focused using an aspheric lens giving a spot size of $\sim 5\mu m$ onto the edge of the sample. The photodetector PD recorded any fluctuations in laser intensity. The suitably polarized light was focussed onto the edge of the GaAs membrane using an aspheric lens (AL) with a focal length of 8 mm. The incoupling was facilitated by back-imaging the laser spot incident on the input facet onto an IR CCD (CAM2). A magnified schematic of the light coupling into the membrane is shown in the inset. The intensity profile of the light propagating in the membrane is captured through the out-of-plane scatter that was imaged using a combination of a 50x long working distance objective (Mitutoyo M Plan Apo SL, 50x/0.42) and a 2x lens adapter onto a SWIR camera (CAM1). The wavelength tuning of the laser and the image acquisition from the SWIR camera was automated using LabVIEW.

Bestfit procedure for $g$:

$$P\left(\frac{I}{\langle I \rangle}\right) = \int_{-i\infty}^{i\infty} K_0\left(2\sqrt{-xI/\langle I \rangle}\right) e^{-g\ln^2\left(\sqrt{1+\frac{x}{g}}+\sqrt{\frac{x}{g}}\right)}$$

The equation describes the distribution of intensity (normalized to the mean) of the speckle pattern formed during the transport of light through the disordered medium. Here, $K_0$ is the modified Bessel function of second kind. The expression is theoretically worked in Ref. [5] and has been experimentally applied for two-dimensional structures in [32], apart from other similar works.

**Acknowledgments:**

SM acknowledges the Swarnajayanti Fellowship from the Department of Science and Technology, Government of India. We acknowledge expert sample fabrication by Monika Emmerling. We acknowledge helpful discussions with P D Garcia and P Lodahl regarding bestfit procedure for the conductance $g$. We thank Sandip Ghosh for lending us the IR laser for these experiments.

**Conflict of interests:**

The authors declare no conflict of interests.

**Contributions:**

R.K. designed the disorder structures and carried out the experiments, and analysed the data. S.Mondal helped in the experimentation and data analysis. The theoretical work was carried out by S.Mondal and M.B. M.K. supervised the fabrication of the GaAs structures. S.Mujumdar conceived and supervised the project and wrote the manuscript. All authors discussed the results and assisted during manuscript preparation.

# Supplementary Information for

# Discrepant transport characteristics under Anderson localization at the two limits of disorder

**This file includes**

- Section S1. Sample Fabrication schematics
- Section S2. Full distribution of $\ln g'$
- Section S3. Density of states
- Section S4. Estimation of the Decay length
- Section S5. Spectral broadening due to finite support.
- Section S6. Mode comparison from experiments and computations.
- Section S7. LDOS and System Size
- Section S8. Theoretically computed $P(\ln g')$
- Section S9. System size dependence of generalized conductance distributions
- Fig. S1. Fabrication technique
- Fig. S2. Experimental conductance distribution
- Fig. S3. Density of states
- Fig. S4. IPR and localization length
- Fig. S5. Computed eigenvalue spectra
- Fig. S6. Mode comparison
- Fig. S7. System size vs LDOS
- Fig. S8. Theoretical conductance distribution
- Fig. S9. Generalized conductance for various system sizes.

**Section S1. Sample Fabrication schematics**

Figure S1 depicts the schematic of the stages of sample fabrication. The process is described in the Methods and Materials Section.

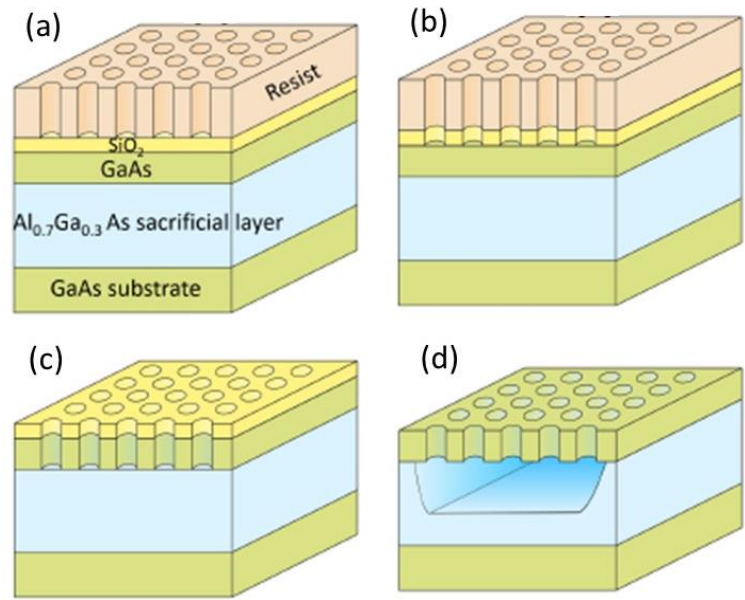

**Fig. S1. Fabrication technique.** Schematics explaining the various stages of sample fabrication.

## Section S2. Full distribution of $\ln g'$

Figure S2 shows the experimentally measured complete distribution $P(\ln g')$ with $g'$, for the various degrees of disorder used. Black curves indicate distributions for passband wavelengths, while red curves depict the same for the bandgap wavelengths. The distributions are asymmetric at weak disorder, while they tend to a normal at strong disorder, implying a lognormal behaviour of $g'$. At weak disorder, a large fraction of the curves is at $\ln g' > 0$, signifying a substantial number of delocalized modes. At strong disorder, however, mostly $\ln g' < 0$.

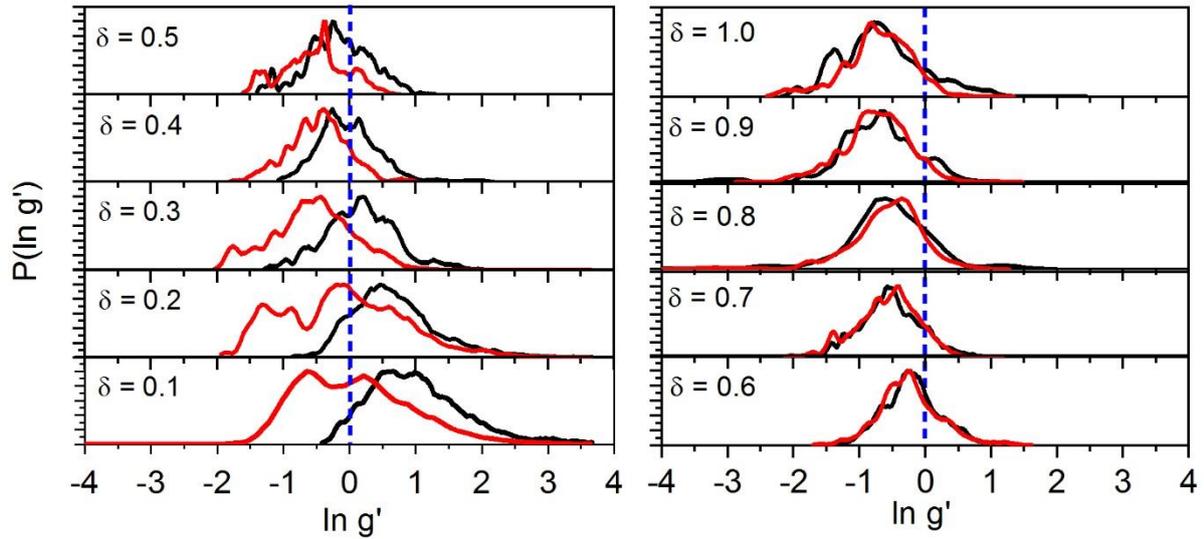

**Fig. S2. Experimental conductance distribution.** Experimentally measured distribution of $\ln g'$ for various disorder strengths. Black curves: Passband wavelengths; Red curves: Bandgap wavelengths.

**Section S3. Density of states**

Figure S3 shows the computed density of states for the triangular lattice of size $20 \times 34$ discussed in the manuscript. The tight-binding Hamiltonian was diagonalized for to obtain the eigenvalues, from which the plot was reconstructed. The lower bandedge at $E = 0.7$ was chosen for analysis.

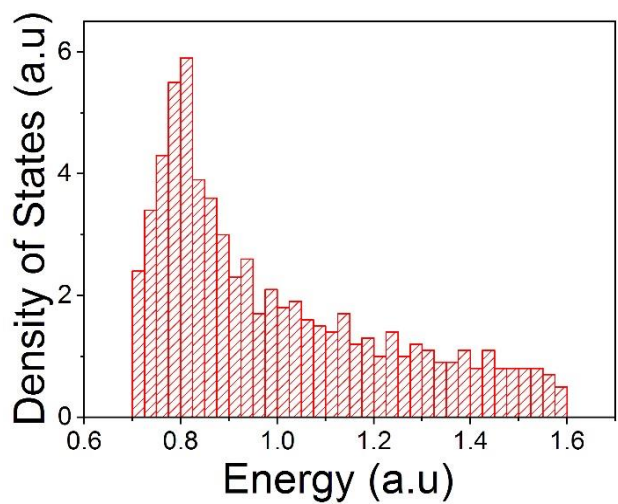

**Fig. S3. Density of states.** Computed density of states from the tight-binding approximation.

**Section S4. Estimation of the Decay length**

The coupling to the boundaries is estimated from the size of the modes, which are quantified by the inverse participation ratios, and their distance from the four boundaries of the sample. As stated in the main text, the loss $\Gamma \propto \exp(-\frac{2R\sqrt{q_2}}{L})$. Here, R is the average distance of the peak of the mode from the boundaries (averaged over the four boundaries), and $q_2$ is the inverse participation ratio of the mode. Figure S4 shows, in black markers, the variation of the decay length with disorder. For comparison, the red markers show the same computed using the localization length that is directly extracted from averaged localized modes. The two profiles are in excellent agreement. However, the localization length cannot be estimated at weak disorder. Our approximation gives a continuous and a monotonic dependence on the disorder strength.

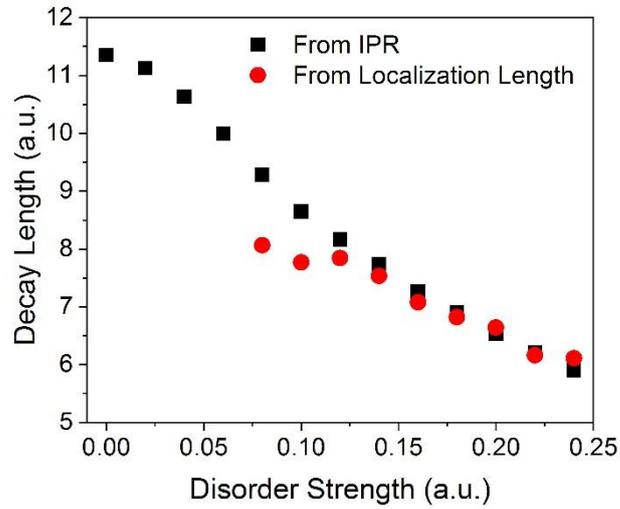

**Fig. S4. IPR and localization length**. Dependence of the decay length (determining the leakage to the boundaries) on disorder strength, as computed from our model (black markers) and from localization length (red markers). The latter technique does not work for weak disorder.

**Section S4. Spectral broadening due to finite support.**

Figure S5 illustrates the spectra due to finite support. For an infinite system, the transmission is in the form of a continuous band. But, for a finite-sized system, the band is discretized into eigenmodes as seen in the top panels. Subplots (a), (b) and (c) show the computed spectra for periodic, weakly disordered and moderately disordered structures. The spectra are shown as a series of δ-functions at the eigenfrequencies. Correspondingly, subplots (d), (e) and (f) show the broadened spectra due to coupling to the surroundings. Analysis of various spectra revealed that the modes are sharper in the stopband, indicating lesser coupling of the stopband modes into the surroundings. Also under strong disorder, the passband modes are narrower than the corresponding passband modes in weak disorder, consistent with the increased localization. Thus, the computed spectra are in agreement with the known behaviour in disordered systems.

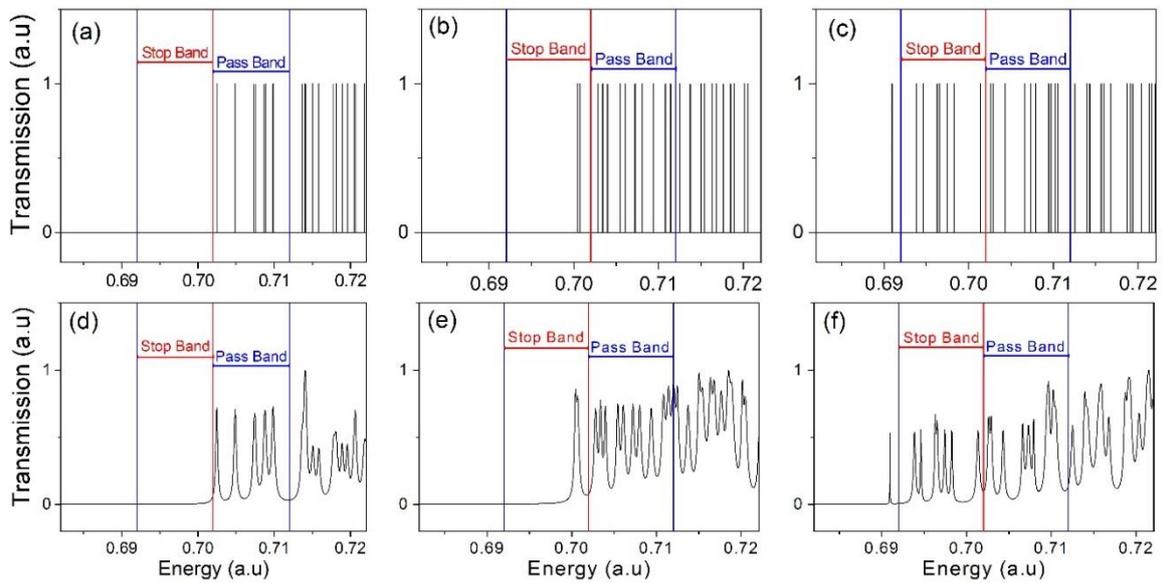

**Fig. S5. Computed eigenvalue spectra.** Computed eigenvalue spectra for (a) periodic (W=0), (b) weakly disordered (W = 0.04) and (c) moderately disordered (W = 0.1) structures. (d), (e), and (f) represent the same spectra after broadening due to finite support, as described in the main manuscript.

**Section S6. Mode comparison from experiments and computations:**

The tight-binding approximation has been extensively used in electronic calculations, but is not generally used for optical experiments. Figure S6 shows the comparison between experimentally measured mode structures at three wavelengths in the periodic structure, with a small inherent fabricational disorder. Here, (a) and (b) show the measurement at two modes in the passband and (c) shows the data at a wavelength in the gap. The modes are affected by the coupling to the surroundings. Subplots (d) and (e) show two mode structures in the passband at minimum disorder, as computed by the model after the finite support was applied. (f) shows a computed intensity distribution in the bandgap, which is in agreement with (c).

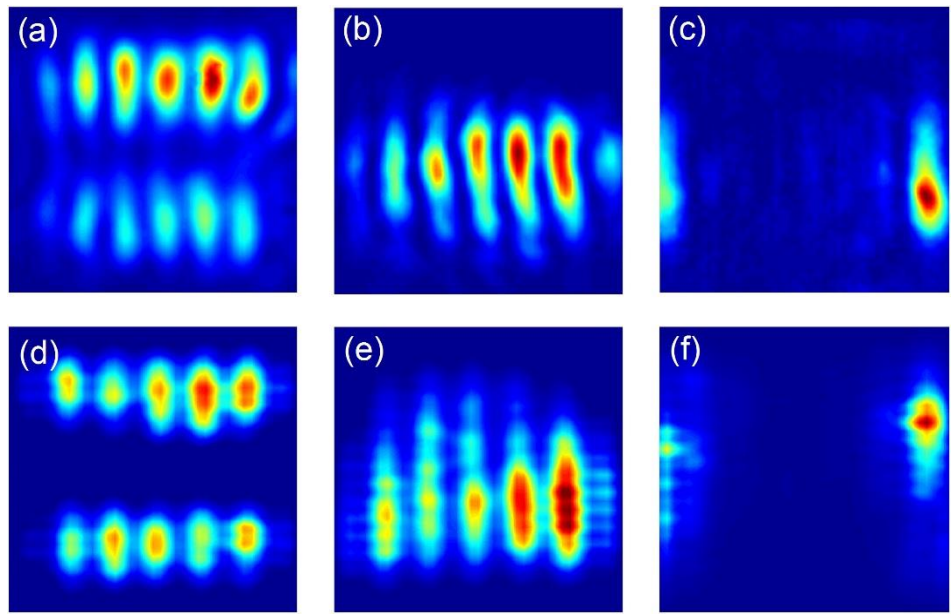

**Fig. S6. Mode comparison.** Top panels show experimentally measured intensity distributions, while bottom panel shows three computed distributions from our model. An excellent agreement is seen between the measured and computed images.

**Section S7. LDOS and System Size**

Figure S7 shows the dependence of local density of states on the system size of a periodic structure, when the open boundaries allow for energy coupling to the environment. The LDOS at the center of the sample is shown, and a clear exponential decay is seen with increasing size.

Our model uses a coupling as defined from the IPR (inverse participation ratio) of the modes, as described in the main paper. The exponential dependence directly translates to a linear dependence of density of states with sample size, which was the main result of PRL, **120,** 237402 (2018).

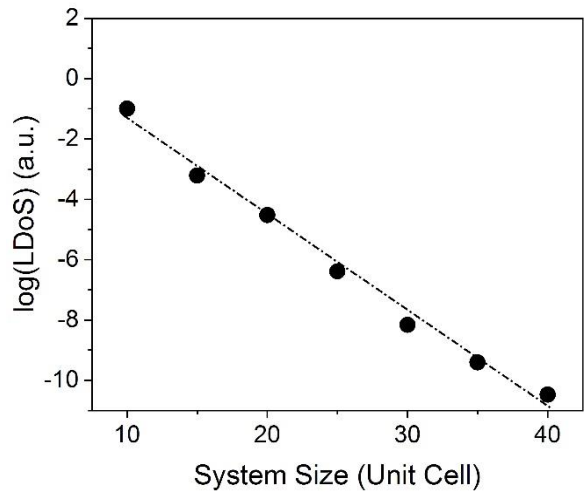

**Fig. S7. System size vs LDOS.** Dependence of local density of states on the system size in a periodic system, when the finite support allows for coupling to the environment.

**Section S8. Theoretically computed $P(\ln g')$:**

Figure S8 shows the theoretically computed distribution $P(\ln g')$ with $g'$, for the various degrees of disorder used. Black curves indicate distributions for passband energies, while red curves depict the same for the bandgap energies. All experimental observations regarding the symmetry of the distributions at strong and weak disorder are excellently reproduced in the model.

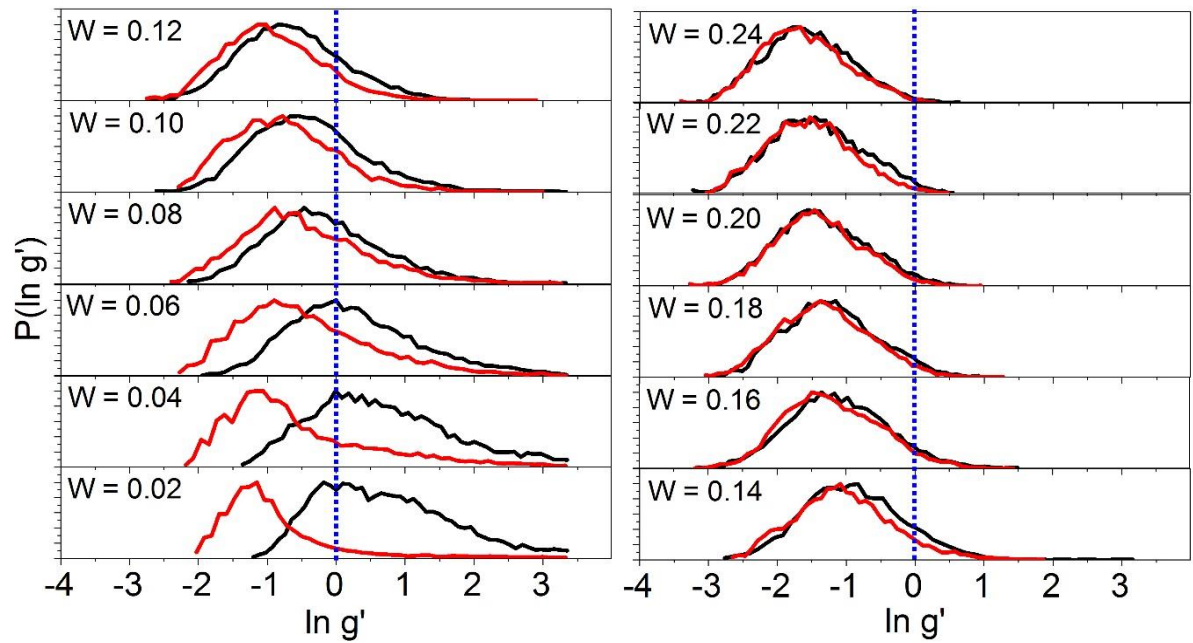

**Fig. S8. Theoretical conductance distribution.** Theoretically calculated distribution of $\ln g'$ for various disorder strengths. Black curves: Passband wavelengths; Red curves: Bandgap wavelengths.

**Section S9. System size dependence of generalized conductance distributions**

The main manuscript illustrates the behaviour of $g'$ for a specific system size of $20 \times 34$. Figure S9 shows the same for various system sizes. Clearly, the mean conductance drops with increasing system size. The qualitative trends, however, remain the same.

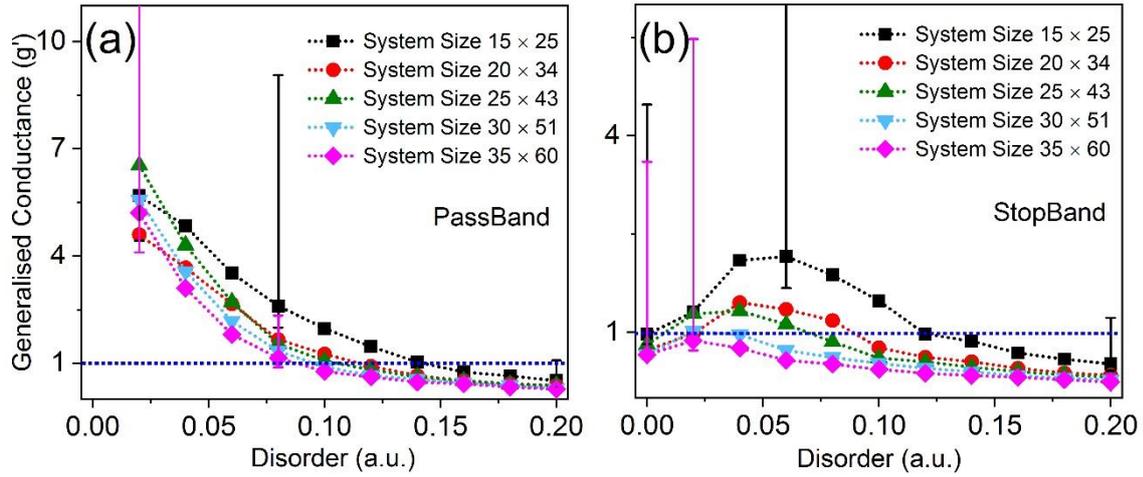

**Fig. S9. Generalized conductance for various system sizes.** As the system size increases, the $g'$ drops. The transition to localization in the passband wavelengths occurs at lower disorder strengths. The trends remain same as discussed in the main manuscript.